\documentclass[aps,prl,preprint,groupedaddress,notitlepage]{revtex4}

\usepackage{graphicx}
\usepackage{wrapfig}

\preprint{NuFact15 - Rio de Janeiro, Brazil - August 2015}

\begin{document}

\baselineskip=16pt


\title{\rightline{\small\rm IIT-CAPP-15-8}Progress on Cherenkov Reconstruction in MICE}
\thanks{\it Presented at NuFact15, 10-15 Aug 2015, Rio de Janeiro, 
Brazil [C15-08-10.2];\\[-0.1in] {\rm work supported by U.S. DOE via the MAP Collaboration.}\\[-0.35in]
}


\author{Daniel M. Kaplan}\email{kaplan@iit.edu}\thanks{presenter}
\author{Michael Drews} \author{Durga Rajaram} \author{Miles Winter}
\affiliation{Illinois Institute of Technology, Chicago, Illinois 60616, USA}
\author{Lucien Cremaldi, 
David Sanders, and Don Summers}
\affiliation{University of Mississippi, Oxford, Mississippi 38677, USA}

\date{\today}

\begin{abstract}
\baselineskip=15pt
Two beamline Cherenkov detectors (Ckov-a,-b) support particle identification in the MICE beamline. Electrons and high-momentum muons and pions can be identified with good efficiency. We report on the Ckov-a,-b performance in detecting pions and muons with MICE Step I data and derive an upper limit on the pion contamination in the standard MICE muon beam.
\end{abstract}


\maketitle

\vspace{-.4in}\section{Introduction}
The international Muon Ionization Cooling Experiment (MICE)\,\cite{MICE} is designed to measure muon ionization cooling\,\cite{Cool}. Cooling is needed
for neutrino factories based on muon decay 
($\mu^- \to e^- \, {\overline{\nu}}_e \, \nu_{\mu}$ \, and \,
$\mu^+ \to e^+ \, \nu_e \, {\overline{\nu}}_{\mu}$)
in  storage rings\,\cite{Factory} and for muon colliders\,\cite{Collider}.

Two high-density aerogel threshold Cherenkov counters\,\cite{Cherenkov}, located just after the first Time of Flight counter (TOF0) in the MICE beamline,
are used in support of muon and pion particle identification. The measured~\cite{MICENote149} refractive indices of the aerogels
in the counters are $n_a=1.069\pm0.003$ in Ckov-a and $n_b=1.112\pm0.004$ in Ckov-b. The corresponding momentum thresholds for muons
(pions) are at  280.5 (367.9) and 217.9 (285.8) MeV/$c$, respectively. Light is collected in each counter
by four 9354KB  eight-inch UV-enhanced phototubes and recorded by CAEN V1731  500\,MS/s flash ADCs (FADCs). 

\vspace{-.1in}\section{Event handling and calibration}
A charge-integration algorithm identifies charge clusters $q_i, i=1$--8 in the FADCs where the ADC value crosses a threshold, marking times $t_1$ and $t_2$ at the threshold crossings, approximating the pulse beginning and end times. The time  $t_{max}$ at the cluster signal maximum is found.  
The charges are converted to a photoelectron count $pe_i$, by subtracting a pedestal $q_{0i}$ and then normalizing by the single photoelectron charge  $q_{1i}$ for each phototube.
For all $q_i > 0$, the total charge, arrival time, $t_1$, and $t_{max}$ are stored per event. 

\begin{figure}[tbp]
\begin{center}
\includegraphics[width=0.495\linewidth, 
trim=15 5 25 15mm,clip]{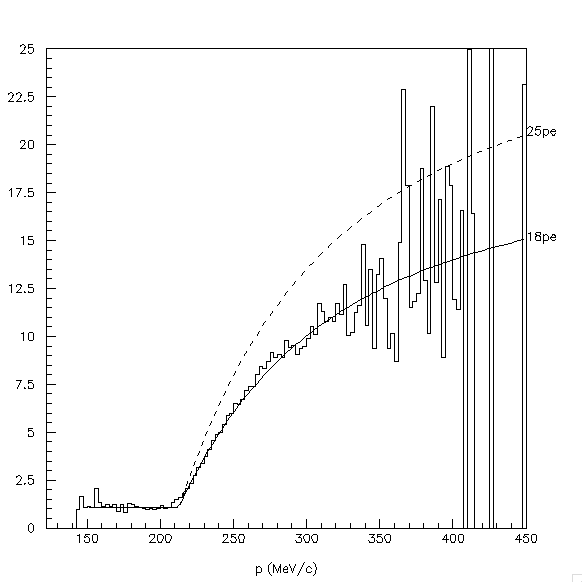}
\includegraphics[width=0.495\linewidth, 
trim=15 5 22 15mm,clip]{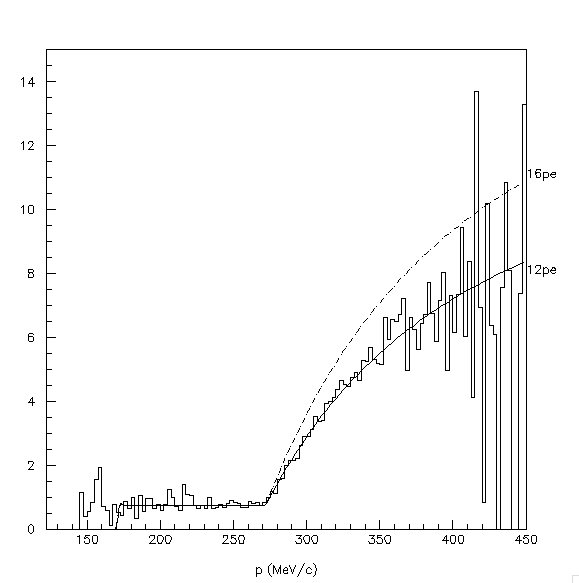}
\caption{\baselineskip=15pt
 Photoelectron ($N_{pe}$) curves versus momentum for muons in (left) Ckov-b  and (right)  Ckov-a.
The $N_{\beta=1}$ values are about 75\% of the values predicted from the
asymptotic photoelectron spectrum of $\beta=1$ electrons (labeled at right)\,---\,not unexpected since for electrons 
TOF0 acts effectively as a ``preshower" radiator.  }
\label{p_thresholds}
\end{center}
\end{figure}

The asymptotic $\beta$=1 light yield  $N_{\beta=1}$ in each counter is measured using the electron peak in MICE calibration-beam runs,  giving 25 and 16 photoelectrons (pe's) in Ckov-b and Ckov-a, respectively, for a nominal run. The photoelectron yields versus momentum are displayed in Fig.~\ref{p_thresholds}.  The observed muon thresholds, $213\pm4$ and $272\pm3$~MeV/$c$, are in reasonable agreement with the expectations  given above. The average number of photoelectrons for normal incidence in the counters can be predicted from the Cherenkov angle
$\cos{\theta_c} = 1/n \beta$, and, near threshold $\beta_{th}=1/n$,
\begin{equation}\label{eq2}
N_{pe} = N_{\beta=1} \times \sin^2{\theta_c}
 = N_{\beta=1} \times (1-(p_{th}/p)^2)\,.
\end{equation}

\begin{wrapfigure}[19]{r}{0.5\textwidth}
\vspace{-.15in}
\centering
\includegraphics[width=\linewidth, 
trim=200 150 235 105 mm,clip]{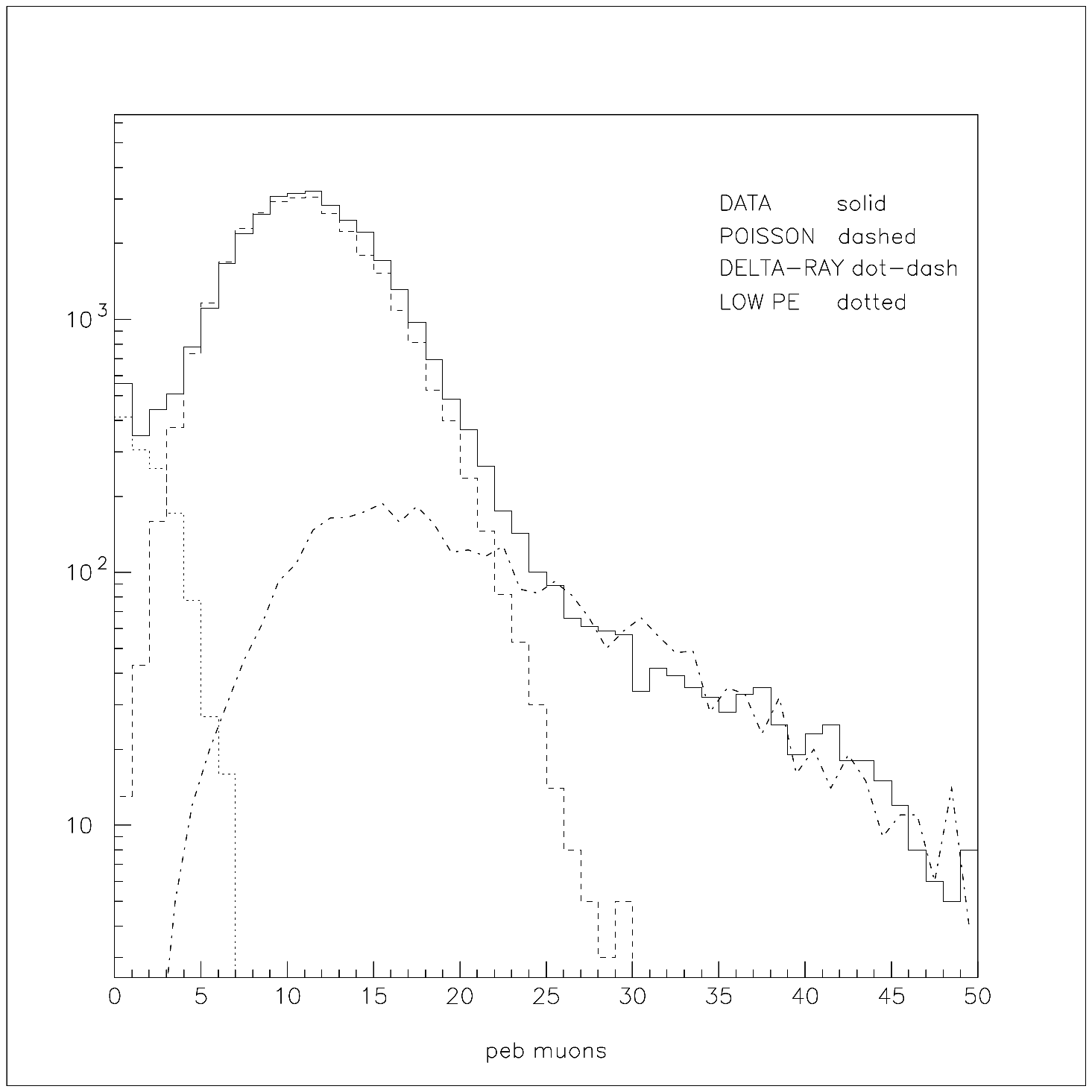}\vspace{-0.15in}
\centerline{\fontfamily{
qhv}\selectfont\tiny Npe}
\caption{\baselineskip=15pt
Typical photoelectron spectrum 
seen for muons or pions above threshold in Ckov-b (solid histogram), together with model fit components:  Poisson (dashed), delta-ray tail (dot-dashed), and anomalous low-$N_{pe}$ component (dotted). }
\label{fig1}
\end{wrapfigure}

As seen in Fig. \ref{fig1}, the photoelectron spectra for 
$\mu$, $\pi$ are observed to be Poisson-like with tails from electromagnetic showers and delta rays produced as the particle traverses TOF0 and the aerogel radiator. Secondary electrons from these processes above about 1 MeV/$c$ produce Cherenkov light 
5--6\% of the time for each particle passage. 
For small-$N_{pe}$ signals, the measured spectra contain more zero-pe events than expected from 
pure Poisson-like behavior $P_0(x) = e^{-x}$, $x=\langle N_{pe}\rangle$. 

\vspace{-.7in}\section{Beam Particle Spectra}
The ``D1'' and ``D2'' dipoles in the MICE beamline~\cite{MICE} predominantly control the beam momentum and particle types transmitted into the MICE spectrometer. In the $p_{tgt} \approx p_{D1} \approx p_{D2}$ setting (calibration mode), the beamline transports a mixture of decay/conversion electrons, decay muons, and primary pions. For $p_{tgt} \approx p_{D1} \approx 0.5 p_{D2}$, backward muon decays from the decay solenoid (DS) are selected. G4beamline\,\cite{G4beamline} Monte Carlo runs indicate  that a small leakage of primary pions through the D2 selection magnet can occur at the $\sim$\,1\% level~\cite{beam-note}. Both these high-momentum pions and their decay muons should be observable  in both Ckov-a and Ckov-b.  Ckov-a can be used effectively to select the high-momentum $\pi,\mu$ events that are just over threshold~\cite{Note}.

\vspace{-.5in}\section{Analysis}
Unambiguous identification of particle species using the Cherenkov detectors (measuring velocity) would require a momentum measurement from the MICE tracker, which was not available in Step I data.  Muons and pions are thus indistinguishable here by the Cherenkov effect. In the  following analysis we look for high-momentum $\pi$ or $\mu$ that trigger Ckov-a. An additional cut on the number of photoelectrons in Ckov-b serves to suppress the $\approx$\,6\% of slow ``background" events that pass the Ckov-a cut due to delta-ray emission.
 
\begin{figure}[tb]
\begin{center}
~\includegraphics[width=0.49\linewidth, height = 3.0 in,trim=20 30 42 15mm,clip]{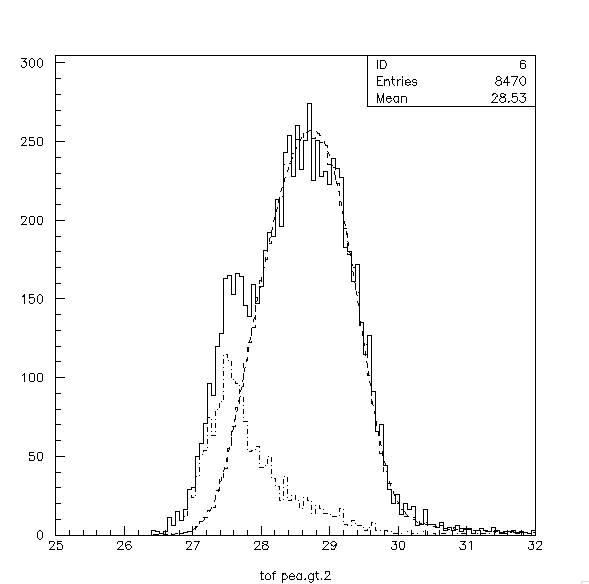}
\includegraphics[width=0.49\linewidth, height = 3.0 in,trim=20 30 42 15mm,clip]{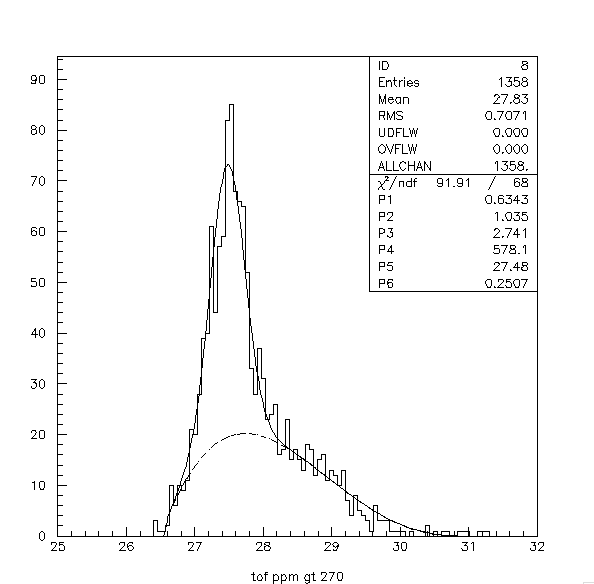}\\[-.1in]
\centerline{\fontfamily{
qhv}\selectfont\tiny ~~~~~~TOF (ns)\hspace{2.875in} TOF (ns)}
\caption{\baselineskip=15pt
Time-of-flight spectrum from TOF0 to TOF1 with (left) pea\,$>2$ cut (solid) and peb\,$>8$ cut (dot-dash), with shape of muon spectrum superimposed (dashed); and  (right) pea\,$>2$ and peb\,$>10$ cuts. 
The peb requirements greatly reduce the delta-ray contribution. Fast $\pi$-$\mu$ are identified as the satellite peak centered at 27.6 ns. }\vspace{-.2in}
\label{fig5}
\end{center}
\end{figure}

We analyzed 120k Step I muon events with $p_{tgt}=400$~MeV/$c$ and $p_{D2}=237$~MeV/$c$ (the ``standard" muon beam settings). We also analyzed 35k muon events with $p_{tgt}=500$~MeV/$c$ and $p_{D2}=294$~MeV/$c$. 
In Fig.~\ref{fig5} we cut away the electron signal (by requiring tof~$>$\,26.4 ns) and also make a Ckov-a $N_{pe}>2$  cut. The shoulder centered at 27.6 ns is made up of fast muons and pions triggering in ${\rm Ckov{\text -}a}$ and  at TOF1. The background events centered approximately at tof\,=\,28 ns are from particles with momenta below threshold in Ckov-a, but giving $N_{pe}>$\,2 Ckov-a light by delta-ray emission. This background is consistent with the expected 6$\%$ contamination level. The tof\,=\,27.6~ns peak corresponds to $p_{\mu}=277$~MeV/$c$ or $p_{\pi}=363$~MeV/$c$, both above threshold in  Ckov-a. 

Fast muons and pions will leave considerable light in  Ckov-b. According to Eq.~\ref{eq2} about 10~pe  will be produced in Ckov-b at $p_{\mu}=270$~MeV/$c$. The probability for simultaneous delta-ray detection in
{\em both} Ckov-a {\em and} Ckov-b will be about $0.06^2=3.6\times10^{-3}$. In Fig.~\ref{fig5} (right) we add a Ckov-b $N_{pe}>\,$10 cut. 
The delta-ray background is substantially reduced to about 500 events. A fit to Gaussian signal and phase-space background of the form ($x\equiv$ time of flight) 
$f = {N({\sqrt{2\pi}}\sigma})^{-1}e^{-(x-\bar{x}^2)/2\sigma^2} + B~ (x-x_{lo})^\alpha (x_{hi}-x)^\beta  $
gives  $539\pm34$ signal events.  When corrected for efficiency~\cite{Note} we obtain $N=1002\pm56$  events. By varying the fitting parameters we find a $\pm101$-event systematic (syst) uncertainty~\cite{Note}.  The fast $\pi$-$\mu$ fraction is thus
$R_{\mu\pi}= ({1002\pm56\pm101})/{118,793}=[0.84\pm0.05\,({\rm stat})\pm0.09\,({\rm syst})]\%$.

If we assume pessimistically that all fast $\pi$-$\mu$ are pions, we can obtain upper limits on the pion fraction: $R_{\mu\pi}<0.97$\%~(90\% CL) and 
$R_{\mu\pi}<1.00\%$~(95\% CL). Any Bayesian model 
would require some prior knowledge of the pion-to-muon ratio in the beam. Estimating this (based on the G4beamline simulation) to be  about 1/20 (or about 50 pions) allows us to estimate the fraction of pions in the beam to be $\pi/\mu \simeq 50/119,000=0.04\%$\,---\,
$\!\!$indeed very small, surpassing the MICE design requirements.
 \vspace{-1.25in}

\subsection{}
\subsubsection{}

\end{document}